\documentclass[12pt,a4paper]{article} 

\usepackage{amsthm}
\usepackage{cancel}
\usepackage{empheq}
\usepackage{float}
\usepackage{blindtext}
\usepackage{enumitem}
\usepackage{enumitem}
\usepackage{amsmath,latexsym,amssymb,amsfonts}
\usepackage{bm}
\usepackage[normalem]{ulem}
\usepackage{bbm}
\usepackage{mathtools,slashed}
\usepackage[hidelinks]{hyperref}
\usepackage{graphicx}
\usepackage[labelfont=bf]{caption}
\usepackage{xcolor}
\usepackage{appendix}
\numberwithin{equation}{section}
\definecolor{magenta}{RGB}{239, 16, 149}
\hypersetup{
 colorlinks,
 linkcolor={blue},
 citecolor={blue},
 urlcolor={blue}
}
\usepackage{bm}
\usepackage{bbold}
\usepackage{mathtools}
\usepackage{braket}

\usepackage[bibstyle=ieee,backend=biber]{biblatex}
\numberwithin{equation}{section}

\usepackage{nicefrac}

\usepackage{tcolorbox}

\newcommand{\Tr}[1]{\mbox{Tr}#1}
\newcommand{\tr}[1]{\mbox{tr}#1}
\newcommand{\diff}{\mathrm{d}}

\begin{document}


\begin{titlepage} 

\begin{center}
{\LARGE \bf The Fierz-Pauli theory on curved spacetime\\[3mm]at one-loop and its counterterms}
\vskip 1.2cm

Leonardo Farolfi$^{\,a}$\footnote{E-mail: leonardo.farolfi2@studio.unibo.it} and Filippo Fecit$^{\,a,b}$\footnote{E-mail: filippo.fecit2@unibo.it}
\vskip 1cm

$^a${\em Dipartimento dFisica e Astronomia ``Augusto Righi", Universit{\`a} di Bologna,\\
via Irnerio 46, I-40126 Bologna, Italy}\\[2mm]

$^b${\em INFN, Sezione di Bologna, via Irnerio 46, I-40126 Bologna, Italy}\\[2mm]

\end{center}
\vskip .8cm

\abstract{We compute the first four heat-kernel coefficients required for the renormalization of Linearized Massive Gravity. We focus on the Fierz-Pauli theory in a curved spacetime, describing the propagation of a massive spin $2$ field in a non-flat background. The background must be on-shell, i.e. must be an Einstein space, in order to ensure a consistent extension of the Fierz-Pauli theory beyond Minkowski space. Starting from the St\"uckelberg formulation with restored gauge invariance, we apply suitable gauge-fixing procedures to simplify the complicated non-minimal kinetic operators of the scalar, vector, and tensor sectors of the theory, reducing them into manageable minimal forms. Using standard techniques, we then compute the Seeley-DeWitt coefficients, with particular emphasis on the fourth coefficient, $a_3(D)$, in arbitrary spacetime dimensions. This result constitutes our main contribution, as it has not been previously reported in full generality in the literature.}

\end{titlepage}

\tableofcontents


\section{Introduction}
The heat-kernel (HK) method is a versatile and powerful tool with applications ranging from black hole entropy to mathematical finance. Its significance in theoretical physics is well established, particularly in the study of effective actions in quantum field theories (QFT) on curved spaces. It is well known that the divergences of the effective action correspond to the so-called Seeley-DeWitt (or heat-kernel) coefficients, which emerge from the short-time expansion of the heat-kernel. Various approaches exist to compute these coefficients, such as iterative solutions to the heat equation \cite{DeWitt:1964mxt, DeWitt:1984sjp, DeWitt:2003pm}, while, from a first-quantized perspective, worldline methods -- which involve a reformulation in terms of relativistic particle path integrals -- have proven to be rather effective, see \cite{Bastianelli:2023oyz, Ahmadiniaz:2024rvi, Manzo:2024gto, Bastianelli:2024vkp, Fecit:2025kqb} for some recent applications in various areas of theoretical physics. \\
In this work, we apply the heat-kernel method to the Fierz-Pauli theory – the theory of a free propagating rank-2 symmetric tensor field $h_{\mu\nu}$ \cite{Fierz:1939zz, Fierz:1939ix} – commonly known as the theory of linearized massive gravity (LMG). This terminology suggests the possibility of interpreting the excitations of $h_{\mu\nu}$ as ``massive gravitons." This simple yet profound idea has sparked extensive research aimed at formulating a modified theory of gravity (see \cite{Hinterbichler:2011tt, deRham:2014zqa, Schmidt-May:2015vnx} for comprehensive reviews), in which the gravitational interaction is mediated by a \textit{massive} spin $2$ particle, in contrast to Einstein's theory, which can be formulated in similar terms but assuming a \textit{massless} particle to begin with. This work aims to analyze the one-loop effective action of a massive spin $2$ field propagating on a curved spacetime background. To achieve this, we first examine how the flat-space action can be consistently extended to curved spacetime while ensuring the absence of ghost-like degrees of freedom, which is a recurring issue in massive gravity theories \cite{Boulware:1972yco}. Building on this framework, we proceed to compute the first four heat-kernel coefficients, with particular emphasis on the fourth coefficient, $a_3(D)$, in arbitrary spacetime dimensions. This coefficient represents our main contribution, as it has not yet been reported in full generality in the literature. Notably, it parametrizes a class of divergences that first appear in dimensions $D\geq 6$.\\
In our endeavor to determine $a_3(D)$, we first have to deal with the challenge of computing the HK coefficients for kinetic differential operators that are non-minimal in the heat-kernel sense: roughly speaking, these are operators with non-trivial higher derivatives structures, as we shall review later. This is often the case with massive field theories. One has then to find a way to reduce the problem in terms of minimal, and manageable, operators. To this aim, we take inspiration from previous works on the subject by Dilkes, Duff, Liu, and Sati \cite{Dilkes:2001av} (see also \cite{Duff:2001zz, Duff:2002sm}). We start by employing the St\"uckelberg formulation of the Fierz-Pauli theory: the so-called “St\"uckelberg trick” \cite{Stueckelberg:1957zz}, which, in a nutshell, consists of introducing redundant variables to restore gauge invariance, proves especially useful in theories where the gauge symmetry gets broken by the presence of a mass term. Restoring gauge symmetry allows us to apply suitable gauge fixings, simplifying the HK computation through standard techniques.\\

The paper is structured as follows. In Section \ref{sec1}, we review the main aspects of the Fierz-Pauli theory and its extension to curved spacetime, including an analysis of its kinetic differential operator. Section \ref{sec3} explores how the introduction of St\"uckelberg fields restores gauge invariance and how to simplify the kinetic operators in order to perform the heat-kernel computations. In Section \ref{sec4}, we compute the heat-kernel coefficients using established results from the literature. Finally, the appendices provide detailed explanations of the computational methods employed.

\section*{Conventions}
We work in a $D$-dimensional Euclidean spacetime endowed with a metric $g_{\mu\nu}(x)$, with Euclidean signature $(+,+,\dots,+)$, that (starting from Sec. \ref{2.3}) satisfies Einstein’s field equations with cosmological constant $\Lambda$
\begin{equation*}
 R_{\mu\nu}-\frac12 g_{\mu\nu}R+g_{\mu\nu}\Lambda=0\ ,
\end{equation*}
which allows to relate the cosmological constant and the Ricci tensor to the Ricci scalar
\begin{equation*}
 \Lambda=\frac{D-2}{2D}R\ ,\quad R_{\mu\nu}=\frac{R}{D} g_{\mu\nu}\ .
\end{equation*}
In other words, we consider a so-called ``Einstein background". The conventions employed for the curvature tensor are:
\begin{equation*}
[\nabla_\mu, \nabla_\nu] V^\lambda = 
R_{\mu\nu}{}^\lambda{}_\rho V^\rho\ , \quad R_{\mu\nu}= R_{\lambda\mu}{}^\lambda{}_\nu\ , \quad R= R^\mu{}_\mu\ , \quad \Omega_{\mu\nu}:=[\nabla_\mu, \nabla_\nu]\ .
\end{equation*}

\section{Fierz-Pauli theory} \label{sec1}
The Fierz-Pauli theory describes the propagation of a massive spin $2$ field carrying 
\begin{equation}
 \# \mathrm{DOFs}=\frac{(D+1)(D-2)}{2}
\end{equation}
degrees of freedom in $D$ spacetime dimensions. Its simplest, second-order, formulation can be expressed in terms of an action principle involving a symmetric and traceful rank-2 tensor field $h_{\mu\nu}(x)$ with a flat spacetime background. Such action is called Fierz-Pauli (FP) theory and has been known since the 1930s \cite{Fierz:1939zz, Fierz:1939ix}
\begin{align} \label{FP}
 S_{\mathcal{FP}}[h_{\mu\nu}]&= \frac12 \int \diff^Dx \, \Bigg[ \underbrace{\partial_\rho h_{\mu\nu} \partial^\rho h^{\mu\nu}-\partial_\mu h \partial^\mu h -2 \partial_\mu h_{\nu\rho}\partial^\nu h^{\mu\rho}+2\partial_\mu h^{\mu\nu} \partial_\nu h}_{=: \, \mathcal{L}_{m=0}[h_{\mu\nu}]}+\underbrace{ m^2\left( h_{\mu\nu} h^{\mu\nu}-h^2\right)}_{=: \, \mathcal{L}_{m}[h_{\mu\nu}]} \Bigg]\ ,
\end{align}
with $h \equiv h^\mu_\mu$. As already mentioned in the introduction, the Fierz-Pauli theory represents the first historical attempt to modify gravity by assuming a massive particle mediating the gravitational interaction. However, one may just as well aside the relation between the FP theory and gravity, and instead interpret the tensor field $h_{\mu\nu}(x)$ as describing massive spin $2$ ``mesons" propagating on either flat or on -- soon to be introduced -- curved background, eventually exploring the interaction with background Einstein gravity (see \cite{Mazuet:2018ysa} for a related discussion). Let us stress that in this work our focus is on the technical computation of the heat-kernel for this theory. The physical interpretation of the theory and its implications for modifications of Einstein gravity remain open to debate, while the potential applications of our results in this context are left for future publications. 

\subsection{Equations of motion and degrees of freedom}
In this subsection, we review some technical aspects that can be inferred starting from \eqref{FP}. The massless action $S_{m=0}[h_{\mu\nu}]:= \tfrac12 \int \diff^Dx \, \mathcal{L}_{m=0}[h_{\mu\nu}] $ is invariant under the gauge symmetry
\begin{equation} \label{gauge}
 \delta h_{\mu\nu}=\partial_\mu \xi_\nu+\partial_\nu \xi_\mu=2\partial_{(\mu}\xi_{\nu)}\ ,
\end{equation}
where $\xi^\mu(x)$ is a spacetime dependent gauge parameter. The latter symmetry is just the linearized version of diffeomorphism invariance: indeed, $ S_{m=0}[h_{\mu\nu}]$ can be either obtained as the most general Lorentz-invariant functional invariant under \eqref{gauge}, or as the linearization of the Einstein-Hilbert action\footnote{We denote the metric tensor in bold in order to hide the spacetime indices, $\mathbf{G}:=G_{\mu\nu}(x)$. Moreover, $G \equiv \det(G_{\mu\nu})$ and $R(\mathbf{G})$ is the Ricci scalar.}
\begin{equation}
 S_{\mathrm{EH}}[\mathbf{G}]=-\frac{1}{2\kappa^2}\int \diff^D x \, \sqrt{G} \big[ R(\mathbf{G})-2\Lambda\big]\ ,
\end{equation}
namely, it can obtained from the latter action by expanding the metric as flat part plus fluctuations:
\begin{equation}
 G_{\mu\nu}(x)=\eta_{\mu\nu}+\kappa h_{\mu\nu}(x)\ ,
\end{equation}
where $\eta_{\mu\nu}$ ($\equiv \delta_{\mu\nu}$ in Euclidean) is the flat metric. The mass term $\mathcal{L}_{m}[h_{\mu\nu}]$ breaks the gauge symmetry, as usual in any gauge field theory. It is actually not the most general mass term that can be added to the invariant action $S_{m=0}[h_{\mu\nu}]$ respecting Lorentz invariance and power counting \cite{Gambuti:2021meo}. However, the choice of the relative $-1$ coefficient between the $h_{\mu\nu} h^{\mu\nu}$ and $h^2$ contractions, known as \emph{Fierz-Pauli tuning}, is the only combination ensuring the propagation of the correct number of degrees of freedom; any other choice would inevitably lead to the presence of a ghost mode with negative energy. To show that the theory describes the dynamics of the correct number of polarizations of a massive spin 2, start from the equations of motion (EOMs) for the tensor field
\begin{equation} \label{eoms}
 \frac{\delta S_{\mathcal{FP}}}{\delta h^{\mu\nu}}=0 \implies \partial^2 h_{\mu\nu} - 2\partial_{(\mu}\partial \cdot h_{\nu)}+ \partial_\mu\partial_\nu h
 + \eta_{\mu\nu}\partial^\rho\partial^\sigma h_{\rho\sigma}
 - \eta_{\mu\nu}\partial^2 h
 =m^2 \left( h_{\mu\nu} - \eta_{\mu\nu} h \right)\ ,
\end{equation}
where the dot denotes contraction over spacetime indices; now, taking the divergence of these equations gives
\begin{align} \label{1.6}
 \underbrace{\partial^\mu \left(\partial^2 h_{\mu\nu} - 2\partial_{(\mu}\partial \cdot h_{\nu)}+ \partial_\mu\partial_\nu h
 + \eta_{\mu\nu}\partial^\rho\partial^\sigma h_{\rho\sigma}
 - \eta_{\mu\nu}\partial^2 h \right)}_{=0}=m^2\partial^\mu \left( h_{\mu\nu} - \eta_{\mu\nu} h \right)\ ,
\end{align}
namely
\begin{equation}
 \partial^\mu h_{\mu\nu} - \partial_\nu h =0\ .
\end{equation}
Finally, plugging the latter equations inside \eqref{eoms} and taking the trace we find
\begin{equation}
 h=0\ .
\end{equation}
Therefore, the equations of motion can be equivalently expressed in terms of the so-called \emph{Fierz-Pauli system}:
\begin{align}
 \left(\partial^2-m^2 \right)h_{\mu\nu}(x)=0\ , \label{KG} \\
 \partial^\mu h_{\mu\nu}(x)=0\ , \label{trans} \\
 h(x)=0\ , \label{trace}
\end{align}
which shows that the tensor field corresponds to a unitary irreducible representation of the Poincar\'e group with mass $m$ and spin $2$. Indeed, \eqref{KG} is the Klein-Gordon equation for the field $h_{\mu\nu}(x)$, while \eqref{trans} and \eqref{trace} represent $D+1$ constraints -- also known as \emph{Fierz-Pauli constraints}, namely transversality and tracelessness -- which reduce the $\tfrac12 D(D+1)$ independent components of $h_{\mu\nu}$ to $\tfrac{1}{2}(D+1)(D-2)$, the DOFs of a massive spin 2. The most fragile among these constraints is undoubtedly the tracelessness one: indeed, violating the FP tuning would lead to the loss of \eqref{trace} already on a flat spacetime, i.e. the simplest formulation of the FP theory, and a ghost-like scalar mode inside $h_{\mu\nu}$ becomes propagating \cite{Hinterbichler:2011tt}. It is not surprising then to learn that taking the theory with a curved spacetime background puts the $h=0$ condition immediately at risk. We shall review it in the next section.

\subsection{Promoting to curved space}
In this work, we consider the propagation of a massive spin $2$ particle on a curved background. Promoting the FP action \eqref{FP} to non-flat spacetime is more challenging than it might appear: the most natural approach\footnote{We will refer to this procedure of coupling a theory to gravity as \emph{minimal} coupling, namely a simple covariant generalization of the Lagrangian, as opposed to the inclusion of \emph{non-minmal} couplings, i.e. enriching the Lagrangian by arbitrary local terms containing powers of the curvature tensor. Note that the minimal procedure gives rise, in the \emph{linear} (massive) theory, to ambiguities because the covariant derivatives do not commute. Without knowledge of a non-linear completion, there is no obvious way of telling which non-minimal couplings to the background curvature to include \cite{Schmidt-May:2015vnx}. The same problem does not occur for (massless) Einstein gravity, where one can safely linearize the Einstein-Hilbert action. Only in recent years, in the case of massive gravity, one can rely on a promising non-linear theory, known as dRGT theory \cite{deRham:2010ik, deRham:2010kj, deRham:2011rn}.} would seemingly be to take the Fierz-Pauli action \eqref{FP} and to promote the ordinary derivatives to full covariant derivatives $\partial\to\nabla$, together with replacing the background metric from flat to curved $\eta_{\mu\nu}\to g_{\mu\nu}$. Such an action would be
\begin{equation} \label{FP'}
 S_{\mathrm{FP'}}[h_{\mu\nu};\mathbf{g}]= \frac12 \int \diff^Dx \sqrt{g} \left[ \nabla_\rho h_{\mu\nu} \nabla^\rho h^{\mu\nu}-\nabla_\mu h \nabla^\mu h -2 \nabla_\mu h_{\nu\rho}\nabla^\nu h^{\mu\rho}+2\nabla_\mu h^{\mu\nu} \nabla_\nu h+m^2\left( h_{\mu\nu} h^{\mu\nu}-h^2\right)\right]\ ,
\end{equation}
where the metric and the covariant derivatives are those of the background $g_{\mu\nu}(x)$. The field equations are just the covariantized version of \eqref{eoms}
\begin{equation}
 \frac{\delta S_{\mathrm{FP'}}}{\delta h^{\mu\nu}}=0 \implies \nabla^2 h_{\mu\nu} - 2\nabla_{\lambda}\nabla_{(\mu} h_{\nu)}{}^{\lambda}+ \nabla_\mu\nabla_\nu h
 + g_{\mu\nu}\nabla^\rho\nabla^\sigma h_{\rho\sigma}
 - g_{\mu\nu}\nabla^2 h
 =m^2 \left( h_{\mu\nu} - g_{\mu\nu} h \right)\ ,
\end{equation}
where $\nabla^2\equiv g^{\mu\nu}\nabla_\mu\nabla_\nu$. One would like to obtain from these equations the covariantized version of the Fierz-Pauli constraints
\begin{align}
 \nabla^\mu h_{\mu\nu}(x)=0\ , \label{1.15} \\
 h(x)=0\ , \label{1.16}
\end{align}
which, for the FP theory,\footnote{It is actually possible to overcome the limitations of the Fierz-Pauli theory and formulate a linear massive gravity theory which consistently propagates a massive spin 2 particle on an \emph{arbitrary} curved background. The starting point is the linearization of the dRGT theory \cite{Bernard:2014bfa, Bernard:2015mkk}.} cannot be achieved on an arbitrary curved background \cite{Bengtsson:1994vn, Hindawi:1995an, Buchbinder:1999ar, Bonifacio:2015rea, Bernard:2017tcg, Fortes:2019asi}. To see that, notice that when saturating the curved EOMs with $\nabla^\mu$, new terms arise because the covariant derivatives do not commute:
\begin{equation} \label{1.17}
 R_{\nu\lambda}\left(\nabla^\lambda h - 2\nabla_\rho h^{\rho\lambda}\right) 
 + \left(\nabla_\rho R_{\nu\lambda} - \nabla_\sigma R^{\sigma}{}_{\rho\lambda\nu}
 \right) h^{\lambda\rho}
 = m^2 \left( \nabla^\mu h_{\mu\nu} - \nabla_\nu h \right)\ ,
\end{equation}
and, differently from \eqref{1.6}, the LHS is not zero. Similarly, taking the trace of the EOMs yields
\begin{equation} \label{1.18}
 (D-2)\nabla^\nu(\nabla^\mu h_{\mu\nu}-\nabla_\nu h)=(1-D) \, m^2h\ .
\end{equation}
It is still possible to find a generalized form of the divergence condition (see the discussion in \cite{Hindawi:1995an}), but not of the tracelessness one; thus, it seems we cannot further reduce the number of propagating DOFs to the ones of a massive spin 2 field. \\

The solution to this problem is well-known \cite{Hinterbichler:2011tt, deRham:2014zqa, Aragone:1971kh, Deser:2006sq}, and the result is that the FP action must\footnote{Note, indeed, that the equations \eqref{1.17}--\eqref{1.18} do not reduce to the curved FP constraints \eqref{1.15}--\eqref{1.16} even on Einstein spacetimes.} be modified and, most importantly, the theory can be consistently formulated only on Einstein spacetimes, for which
\begin{equation} \label{E}
 R_{\mu\nu}= \lambda g_{\mu\nu}\ .
\end{equation}
The upshot is that the curved Fierz-Pauli constraints can be derived from an action, whose form is given by covariantizing the linearized Einstein-Hilbert action plus the FP mass term and, in addition, including an extra term proportional to the scalar curvature. A detailed derivation of the Fierz-Pauli Lagrangian on curved Einstein space can be found in \cite{Buchbinder:1999ar}, where the authors begin their analysis by considering all possible non-minimal terms involving the curvature tensor, each multiplied by a dimensionless coefficient $\alpha_j$ in front of them:
\begin{align}
 S_{\mathrm{FP''}}[h_{\mu\nu};\mathbf{g}]&=S_{\mathrm{FP'}}[h_{\mu\nu};\mathbf{g}] \\
 &\phantom{=}+ \int \diff^Dx \sqrt{g} \left[ \alpha_1 \, R \, h_{\mu\nu} h^{\mu\nu}
+\alpha_2 \, R \, h^2
+\alpha_3 \, R^{\lambda\mu\rho\nu} h_{\lambda\rho} h_{\mu\nu}
+\alpha_4 \, R^{\mu\nu} h_{\mu\sigma} h_\nu{}^\sigma
+\alpha_5 \, R^{\mu\nu} h_{\mu\nu} h
 \right]\ , \nonumber
\end{align}
which is the most general action for massive spin 2 field in curved spacetime quadratic in derivatives and consistent with the flat limit. By studying the resulting equations of motion and the relative constraints, they demonstrated that only a restricted subset of all possible non-minimal couplings to the curvature has to be included, ultimately arriving at the final action
\begin{equation} \label{FPcurved}
S_{\mathrm{FP}}[h_{\mu\nu};\mathbf{g}]= \frac12 \int \diff^Dx \sqrt{g} \Big[ \mathcal{L}_{m=0}[h_{\mu\nu};\mathbf{g}]+\mathcal{L}_{m}[h_{\mu\nu};\mathbf{g}] \Big]
\end{equation}
where the mass term remains unchanged, while the massless Lagrangian now reads
\begin{equation} 
 \mathcal{L}_{m=0}[h_{\mu\nu};\mathbf{g}] = \nabla_\rho h_{\mu\nu} \nabla^\rho h^{\mu\nu}-\nabla_\mu h \nabla^\mu h -2 \nabla_\mu h_{\nu\rho}\nabla^\nu h^{\mu\rho}+2\nabla_\mu h^{\mu\nu} \nabla_\nu h-\frac{2R}{D}\left( h_{\mu\nu} h^{\mu\nu}-\frac12 h^2\right)\ .
\end{equation}
Taking the equations of motion $E_{\mu\nu}=0$, with $E_{\mu\nu}:=\frac{\delta S_{\mathrm{FP}}}{\delta h^{\mu\nu}}$, that now read
\begin{equation} \label{eomscurved}
E_{\mu\nu}= \nabla^2 h_{\mu\nu} - 2\nabla_{\lambda}\nabla_{(\mu} h_{\nu)}{}^{\lambda}+ \nabla_\mu\nabla_\nu h
 + g_{\mu\nu}\nabla^\rho\nabla^\sigma h_{\rho\sigma}
 - g_{\mu\nu}\nabla^2 h+\frac{2R}{D}\left( h_{\mu\nu}- \frac12 g_{\mu\nu}h \right)
 -m^2 \left( h_{\mu\nu} - g_{\mu\nu} h \right)\ ,
\end{equation}
and saturating with $\nabla^\mu$, we get 
\begin{equation}
 \nabla^\mu h_{\mu\nu} - \nabla_\nu h =0\ ,
\end{equation}
having exploited the following Einstein space simplifications:
\begin{align}
 \nabla_\nu R=0 \; \to \; \nabla_\rho R_{\mu\nu}=0 \; \to \; \nabla^\mu R_{\mu\nu\rho\sigma}=0\ ,
\end{align}
obtained by combining \eqref{E} with the second Bianchi identity $\nabla_{[\lambda}R_{\rho\sigma]\mu\nu}=0$, for $D>2$. Finally, taking $\nabla^\mu\nabla^\nu E_{\mu\nu}=0$ and plugging the result inside $g^{\mu\nu}E_{\mu\nu}=0$, we obtain
\begin{equation}
 \left( m^2 -\frac{D-2}{D(D-1)}R \right) \, h=0\ ,
\end{equation}
from which\footnote{Notice that the particle content changes depending on the value of $m^2$. The theory describes the expected degrees of freedom of a massive spin 2 as long as $m^2 \neq \tfrac{D-2}{D(D-1)}R$, which is the so-called Higuchi bound \cite{Higuchi:1986py}. If the mass assumes that boundary value, then the theory acquires a scalar gauge symmetry and is known as ``partially massless", as it propagates one fewer degree of freedom than usual \cite{Deser:1983mm}.} we get the curved tracelessness constraint \eqref{1.16} and consequently also the curved transversality one \eqref{1.15}. Plugging them inside \eqref{eomscurved}, we get the curved-space Klein-Gordon equation: 
\begin{equation} \label{1.20}
 (\nabla^2 -m^2) \, h_{\mu\nu}(x) -2 R_{\rho(\mu\nu)\lambda} \, h^{\rho\lambda}(x)=0\ .
\end{equation}
Therefore, the equations of motion can be equivalently written in terms of the curved Fierz-Pauli system \eqref{1.15}, \eqref{1.16} and \eqref{1.20}, which highlights the propagation of the correct number of degrees of freedom for a massive spin 2, without ghosts, on curved Einstein background. In conclusion, note that the massless part of \eqref{FPcurved} is invariant under the gauge symmetry
\begin{equation} \label{gaugecurvo}
 \delta h_{\mu\nu}=2\nabla_{(\mu}\xi_{\nu)}\ ,
\end{equation}
whereas the total action is not, due to the inclusion of the FP mass term. 

\subsection{Kinteic operator and Heat-Kernel} \label{2.3}
The FP theory on curved Einstein spacetime can be recast in the following convenient form\footnote{We will always assume that the fields decay quickly enough at infinity when integrating by parts, hence omitting boundary terms.}
\begin{equation}
 S_{\mathrm{FP}}[h_{\mu\nu};\mathbf{g}]= \frac12 \int \diff^Dx \sqrt{g} \, h_{\mu\nu} \, \Delta^{\mu\nu\rho\sigma}_{(\mathrm{II})} \, h_{\rho\sigma} 
\end{equation}
in terms of the differential operator
\begin{align} \label{Delta2}
 \Delta^{\mu\nu\rho\sigma}_{(\mathrm{II})}=\left(-\nabla^2 +m^2\right) \left( g^{\mu(\rho}g^{\sigma)\nu} -g^{\mu\nu} g^{\rho\sigma}\right)+2R^{\mu(\rho\sigma)\nu}+\frac{R}{D}g^{\mu\nu}g^{\rho\sigma}-2g^{\rho\sigma}\nabla^{(\mu}\nabla^{\nu)}+2g^{(\mu(\rho}\nabla^{\nu)}\nabla^{\sigma)}\ ,
\end{align}
We aim to compute, perturbatively, the one-loop (Euclidean) effective action $\Gamma_{\mathrm{E}}[\mathbf{g}]$, defined in the path integral formulation as
\begin{equation}
 e^{-\Gamma_{\mathrm{E}}[\mathbf{g}]}= \int [\mathcal{D}\mathbf{h}] \, e^{-S_{\mathrm{FP}}[h_{\mu\nu};\mathbf{g}]}\ .
\end{equation}
As the action is quadratic, the result can be expressed in terms of the functional determinant of the $\Delta_{(\mathrm{II})}$ operator \cite{Avramidi:2000bm}
\begin{equation}
 \Gamma_{\mathrm{E}}[\mathbf{g}]=\frac12 \log\mathrm{Det}\big[ \Delta_{(\mathrm{II})} \big]= \frac12 \Tr \log\big[ \Delta_{(\mathrm{II})} \big] ,
\end{equation}
where $\Tr(\dots)\equiv\int_{\mathcal{M}} \diff^Dx\sqrt{g}\,\tr(\dots)$ denotes the functional trace. It is not an easy task to compute said determinant directly, even though at the perturbative level things become more manageable. One possibility consists of evaluating the related heat-kernel expansion: indeed, using the Schwinger-DeWitt proper-time parametrization we can express the effective action in terms of an integral over Euclidean time $\beta$ of the trace of the heat-kernel $K(x,x';\beta)$ -- defined via
\begin{equation} \label{heat}
 \Tr \, e^{-\beta \Delta_{(\mathrm{II})}}= \int \diff^Dx \, \sqrt{g} \; \tr \, K(x,x;\beta)
\end{equation}
-- at coinciding point $x' \to x$, as
\begin{equation} \label{eff}
 \Gamma_{\mathrm{E}}[\mathbf{g}]=-\frac12 \int \diff^Dx \, \sqrt{g} \int_{0}^{+\infty} \frac{\diff\beta}{\beta} \, \tr \, K(x,x;\beta)\ ,
\end{equation}
where the trace in the RHS of \eqref{heat} is to be performed over spacetime indices carried by the fields on which $\Delta_{(\mathrm{II})}$ acts upon. The perturbative evaluation of \eqref{eff} is studied in the limit $\beta \to 0$, in which the diagonal heat-kernel can be expressed in terms of local functions of the curvature invariants: these are the so-called heat-kernel, or Seeley-DeWitt (SdW), coefficients $a_j(x,D)$ related to the operator $\Delta_{(\mathrm{II})}$ as \cite{DeWitt:1964mxt, DeWitt:1984sjp, DeWitt:2003pm}
\begin{equation} \label{exp}
 (4\pi \beta)^{\nicefrac{D}{2}} \, K(x,x;\beta) \, \sim \, a_0(D) + a_1(x,D) \, \beta + a_2(x,D) \, \beta^2+ a_3(x,D) \ \beta^3 + \dots \quad \text{as} \quad \beta\to0\ ,
\end{equation}
where $a_0(D)$ counts the number of degrees of freedom. Following the Schwinger-DeWitt method \cite{Vassilevich:2003xt}, these coefficients can be expressed in terms of the metric and gauge invariants of the manifold, as reviewed in appendix \ref{appB}. Our aim is to compute the first four coefficients, with the fourth one $a_3(x,D)$ not fully known in the literature. However, things become slightly more complicated when dealing with massive theories. \\

In principle there is no straightforward heat-kernel expansion of the form \eqref{exp} when the operator is \emph{non-minimal}. In the heat-kernel context,\footnote{Not to be confused with the meaning that ``non-minimal" assumes regarding the coupling to gravity.} we call a (second-order) differential operator non-minimal when the part of the operator with the highest derivatives -- the principal symbol \cite{Vassilevich:2003xt} -- has a non-trivial matrix structure. For instance, an operator of Laplace type
\begin{equation}
 \Delta=-g^{\mu\nu}\nabla_\mu\nabla_\nu+V\ ,
\end{equation}
with $V$ a matrix potential, is a \emph{minimal} operator, as the principal symbol is the metric. A \emph{non-minimal} operator presents a more general principal symbol: for instance, the kinetic operator of the Proca field on curved spacetime has the form \cite{Buchbinder:2017zaa, Ruf:2018vzq}
\begin{equation}
\Delta^{\mu\nu}=(-\nabla^2+m^2) \,g^{\mu\nu}+\nabla^\mu \nabla^\nu +R^{\mu\nu}\ ,
\end{equation}
and the offending term $\nabla^\mu \nabla^\nu$ is responsible for the complicated structure of the leading derivative terms. This second scenario is the case we are dealing with, due to the $-2g^{\rho\sigma}\nabla^{(\mu}\nabla^{\nu)}+2g^{(\mu(\rho}\nabla^{\nu)}\nabla^{\sigma)}$ terms in the LMG operator \eqref{Delta2}. Thus we have to find a way to reduce the kinetic operator for the tensor field in minimal form $\hat{\Delta}_{(\mathrm{II})}$, where\footnote{Throughout the work, we will always denote \emph{minimal} operators with the hat $\hat{\Delta}$, as opposed to their \emph{non-minimal} versions $\Delta$.}
\begin{equation}
 \Delta^{\mu\nu\rho\sigma}_{(\mathrm{II})} =\hat{\Delta}^{\mu\nu\rho\sigma}_{(\mathrm{II})}-2g^{\rho\sigma}\nabla^{(\mu}\nabla^{\nu)}+2g^{(\mu(\rho}\nabla^{\nu)}\nabla^{\sigma)}\ ,
\end{equation}
namely
\begin{equation} \label{hatDelta2}
\hat{\Delta}^{\mu\nu\rho\sigma}_{(\mathrm{II})}=\left(-\nabla^2 +m^2\right) \left( g^{\mu(\rho}g^{\sigma)\nu} -g^{\mu\nu} g^{\rho\sigma}\right)+2R^{\mu(\rho\sigma)\nu}+\frac{R}{D}g^{\mu\nu}g^{\rho\sigma}\ .
\end{equation}
There exist different viable approaches to perform the reduction to minimal form, the most prominent one probably being the generalization of the Schwinger-DeWitt algorithm developed by Barvinsky and Vilkovisky \cite{Barvinsky:1985an}. Here we find more suitable to our case to follow a path integral-based approach, akin to what has been previously done in \cite{Dilkes:2001av}. This procedure will be covered in the next section. \\
 
Before concluding this section, let us define a slightly different version of the spin 2 operator, for notational simplicity of the upcoming calculations: the operator
\begin{equation} \label{schifo}
\tilde{\Delta}^{\chi \;\mu\nu\rho\sigma}_{(\mathrm{II})}=-\nabla^2 \left( g^{\mu(\rho}g^{\sigma)\nu} -\chi \, g^{\mu\nu} g^{\rho\sigma}\right)+m^2\left( g^{\mu(\rho}g^{\sigma)\nu} -\chi' \, g^{\mu\nu} g^{\rho\sigma}\right)+2R^{\mu(\rho\sigma)\nu}+\frac{R}{D}g^{\mu\nu}g^{\rho\sigma}\ ,
\end{equation}
is still a minimal operator, but, due to the presence of the parameters $\chi$ and $\chi'$, it may treat the traceless and scalar part of the spin 2 field differently.

\section{Gauge-fixing manipulations} \label{sec3}
The idea is to find suitable gauge choices that reduce the operator to its minimal form. To achieve this, we first have to reintroduce the broken symmetry, through the St\"uckelberg formulation of the FP theory. \\

We introduce the St\"uckelberg vector field $A_\mu(x)$ through a shift patterned after \eqref{gaugecurvo}, i.e. \cite{Stueckelberg:1957zz}
\begin{equation}
 h_{\mu\nu} \to h_{\mu\nu}+\frac{1}{m} \left( \nabla_\mu A_\nu+\nabla_\nu A_\mu \right)\ .
\end{equation}
While the massless action remains unchanged, from the mass term we get new contributions
\begin{equation}
 \mathcal{L}_{\mathrm{m}}[h_{\mu\nu};\mathbf{g}]\to \mathcal{L}_{\mathrm{m}}[h_{\mu\nu},A_\mu;\mathbf{g}]= \mathcal{L}_{\mathrm{m}}[h_{\mu\nu};\mathbf{g}] + F_{\mu\nu}F^{\mu\nu}-\frac{4R}{D}A_\mu A^\mu +4m\left( h_{\mu\nu}\nabla^\mu A^\nu -h\nabla_\mu A^\mu\right)\ ,
\end{equation}
where $F_{\mu\nu}=\nabla_\mu A_\nu-\nabla_\nu A_\mu$. The total action is now invariant under
\begin{equation}
 \delta h_{\mu\nu}=2\nabla_{(\mu}\xi_{\nu)}\ , \quad \delta A_\mu=-m \, \xi_\mu\ .
\end{equation}
We chose to gauge-fix this redundancy with
\begin{equation} \label{gauge-fix1}
 \nabla^\nu h_{\mu\nu}-\frac12 \nabla_\mu h+mA_\mu=0\ ,
\end{equation}
which has the effect of simplifying the non-minimal terms inside $\Delta_{(\mathrm{II})}$ together with canceling the mixed $h$ -- $A_\mu$ term above. The total action reads now
\begin{equation} \label{2.5}
 S_{\mathrm{FP}}[h_{\mu\nu},A_\mu;\mathbf{g}]= \frac12 \int \diff^Dx \sqrt{g} \left[ h_{\mu\nu} \, \tilde{\Delta}^{\chi \;\mu\nu\rho\sigma}_{(\mathrm{II})} \, h_{\rho\sigma} 
+F_{\mu\nu}F^{\mu\nu}+\mathfrak{m}^2 A_\mu A^\mu-2m h \nabla_\mu A^\mu \right]\ ,
\end{equation}
having introduced the following notation for the mass-like term of the vector field
\begin{equation}
 \mathcal{L}_{\mathfrak{m}}[A_{\mu};\mathbf{g}]=\mathfrak{m}^2A_\mu A^\mu
\end{equation}
with
\begin{equation} \label{mass}
 \mathfrak{m}^2=2m^2-\frac{4R}{D}\ .
\end{equation}
The tensor field operator is now minimal and is given by \eqref{schifo} with $\chi=\nicefrac12$ and $\chi'=1$. Note, however, the presence of a resilient mixed tensor--vector term in the action: to take care of that, it is enough to make the following change of variables
\begin{equation} \label{rep}
 A_\mu \to A_\mu +\frac{mD}{4R-2m^2D} \, \nabla_\mu h\ ,
\end{equation}
so that the action becomes diagonal:
\begin{equation} \label{parz}
 S_{\mathrm{FP}}[h_{\mu\nu},A_\mu;\mathbf{g}]= \frac12 \int \diff^Dx \sqrt{g} \left[ h_{\mu\nu} \, \tilde{\Delta}^{\chi \;\mu\nu\rho\sigma}_{(\mathrm{II})} \, h_{\rho\sigma} 
+F_{\mu\nu}F^{\mu\nu}+\mathfrak{m}^2 A_\mu A^\mu \right]\ .
\end{equation}
The price that has been paid is that now the tensor field kinetic operator gets slightly more complicated to handle, being the one previously defined in \eqref{schifo} with $\chi'=1$, and with now
\begin{equation}
 \chi=\frac{m^2D-R}{m^2D-2R}\ .
\end{equation}
Let us comment on the fact that the replacement \eqref{rep} is not well-defined for $m^2=\frac{2R}{D}$. This suggests that it is not always possible to use it and one should find another way to proceed with the computation: we illustrate this alternative path in appendix \ref{appA}. For the time being, we will assume that $m^2\neq\frac{2R}{D}$, as the change of variables \eqref{rep}, although not crucial, drastically simplifies the calculation. \\

In principle, one could use the resulting action \eqref{parz} to perform the path integral, as it would result in two distinct functional integrations over a massive spin 2 field and a massive spin 1 field, without forgetting to include the Faddeev-Popov ($\Phi\Pi$) determinant associated with the gauge choice \eqref{gauge-fix1}. However, it is not wise to deal with the heat-kernel expansion of tensor and vector kinetic operators as they stand: for instance, one may notice that the vector one is \emph{not} a minimal operator. It is preferable to find a way to further simplify their expressions, by means of another gauge-fixing manipulation. \\

Let us introduce a second St\"uckelberg field, a scalar $\varphi(x)$, by 
\begin{equation}
 A_\mu \to A_\mu +\frac{1}{m} \nabla_\mu \varphi\ .
\end{equation}
It is clear that the only term that is not left invariant is the mass-like term for the spin one. Indeed,
\begin{equation}
 \mathcal{L}_{\mathfrak{m}}[A_{\mu};\mathbf{g}] \to \mathcal{L}_{\mathfrak{m}}[A_{\mu},\varphi;\mathbf{g}]=\mathcal{L}_{\mathfrak{m}}[A_{\mu};\mathbf{g}]+\frac{2\mathfrak{m}^2}{m}A^\mu\nabla_\mu \varphi +\frac{\mathfrak{m}^2}{m^2}\nabla_\mu \varphi\nabla^\mu\varphi\ .
\end{equation}
The total action now has an additional gauge symmetry:
\begin{equation} \label{gauge2}
 \delta A_\mu=\nabla_\mu \Sigma\ , \quad \delta\varphi=-m\Sigma\ ,
\end{equation}
with $\Sigma(x)$ a scalar parameter. We choose the following gauge-fixing condition
\begin{equation} \label{gauge-fix2}
 \nabla^\mu A_\mu+\frac{\mathfrak{m}^2}{2m}\varphi=0\ ,
\end{equation}
which makes the vector kinetic operator minimal, so that the total action reads
\begin{equation} \label{tot}
 S_{\mathrm{FP}}[h_{\mu\nu},A_\mu,\varphi;\mathbf{g}]= \frac12 \int \diff^Dx \sqrt{g} \left[ h_{\mu\nu} \, \tilde{\Delta}^{\chi \;\mu\nu\rho\sigma}_{(\mathrm{II})} \, h_{\rho\sigma} 
+ 2A_{\mu} \, \hat{\Delta}^{\mu\nu}_{(1)} \, A_{\nu} +\frac{\mathfrak{m}^2}{m^2} \, \varphi \, \hat{\Delta}_{(0)} \, \varphi \right]\ ,
\end{equation}
up to boundary terms. It is clearly in diagonal form, with the tensor, vector, and scalar sector totally unmixed and written in terms of their respective differential operators, with
\begin{equation} \label{Delta1}
 \hat{\Delta}^{\mu\nu}_{(1)}=\left(-\nabla^2 +m^2-\frac{R}{D}\right) \, g^{\mu\nu}\ ,
\end{equation}
and with
\begin{equation} \label{Delta0}
 \hat{\Delta}_{(0)}=-\nabla^2 +m^2-\frac{2R}{D}\ .
\end{equation}
The resulting partition function reads in full glory
\begin{equation}
 Z[\mathbf{g}] = \int [\mathcal{D}\mathbf{h}] \, [\mathcal{D}\mathbf{A}] \, [\mathcal{D}\varphi] \, e^{-S_{\mathrm{FP}}[h_{\mu\nu},A_\mu,\varphi;\mathbf{g}]} \, \mathrm{Det}\big[ \hat{\Delta}_{(1)} \big] \, \mathrm{Det}\big[ \hat{\Delta}_{(0)} \big]\ ,
\end{equation}
where we have included the $\Phi\Pi$ determinants $\mathrm{Det}\big[ \hat{\Delta}_{(1)} \big]$ and $\mathrm{Det}\big[ \hat{\Delta}_{(0)} \big]$, that must be inserted inside the path integral following the Faddeev-Popov procedure, stemming from the vector \eqref{gauge-fix1} and scalar \eqref{gauge-fix2} gauge-fixings, respectively. Let us stress the structure of the total action by rewriting it as
\begin{equation} \label{2.19}
 S_{\mathrm{FP}}[h_{\mu\nu},A_\mu,\varphi;\mathbf{g}]=\frac12S_{\mathrm{gra}}[h_{\mu\nu};\mathbf{g}]+S_{\mathrm{vec}}[A_\mu;\mathbf{g}]+\frac{\mathfrak{m}^2}{2m^2} S_{\mathrm{sca}}[\varphi;\mathbf{g}]\ ,
\end{equation}
where the definition of each contribution can be easily deduced from \eqref{tot}. Note that we collected the prefactors in order to recast each kinetic operator with a standard principal symbol, i.e. $\Delta = -\nabla^2+\dots$. \\

The final expression is still complicated to exploit for heat-kernel calculation. In particular, the problematics originate from the spin 2 kinetic operator \eqref{schifo}, which treats asymmetrically the scalar and the traceless part of the tensor field. Although it is likely possible to use it as it is, we would like to recast it in order to exploit the calculations of \cite{Bastianelli:2023oca}. To that aim, let us decompose $h_{\mu\nu}$ into traceless and scalar part
\begin{equation} \label{decompose}
 h_{\mu\nu}=\Phi_{\mu\nu}+\frac{1}{D}g_{\mu\nu} \, \phi\ ,
\end{equation}
with
\begin{equation}
 g_{\mu\nu} \Phi^{\mu\nu}=0\ , \quad \phi:=g_{\mu\nu}h^{\mu\nu}\ .
\end{equation}
The spin 2 field action splits as
\begin{equation}
 S_{\mathrm{gra}}[h_{\mu\nu};\mathbf{g}]=S_{\mathrm{gra}}^{(\mathrm{I})}[\Phi_{\mu\nu};\mathbf{g}]+\mathcal{X}\,S_{\mathrm{gra}}^{(\mathrm{II})}[\phi;\mathbf{g}]\ ,
\end{equation}
where
\begin{align}
 S_{\mathrm{gra}}^{(\mathrm{I})}[\Phi_{\mu\nu};\mathbf{g}]&= \int \diff^Dx \sqrt{g} \left[ \Phi_{\mu\nu} \, \hat{\Delta}^{\mu\nu\rho\sigma}_{(2)} \, \Phi_{\rho\sigma} \right]\ , \label{2.23} \\
 S_{\mathrm{gra}}^{(\mathrm{II})}[\phi;\mathbf{g}]&=\int \diff^Dx \sqrt{g} \left[ \phi \, \hat{\Delta}_{(0)} \, \phi \right]\ ,
\end{align}
with $\hat{\Delta}_{(0)}$ previously defined in \eqref{Delta0} and where we denoted $\hat{\Delta}^{\mu\nu\rho\sigma}_{(2)}$ the following minimal differential operator
\begin{equation} \label{DeltaI}
 \hat{\Delta}^{\mu\nu\rho\sigma}_{(2)}=\left(-\nabla^2 +m^2\right) \, g^{\mu(\rho}g^{\sigma)\nu} +2R^{\mu(\rho\sigma)\nu}\ ,
\end{equation}
and where we gathered together the following prefactors
\begin{equation}
 \mathcal{X}=\frac{m^2 D (D-1)-(D-2)R}{D(m^2 D-2R)}\ .
\end{equation}
Note that the operator \eqref{DeltaI} is acting on the rank-2 \emph{traceless} tensors sector.

\section{Extracting the Heat-Kernel coefficients} \label{sec4}
The partition function can be finally recast as
\begin{equation}
 Z[\mathbf{g}] = \int [\mathcal{D}\mathbf{\Phi}] \, [\mathcal{D}\mathbf{A}] \, [\mathcal{D}\varphi] \, [\mathcal{D}\phi] \, e^{-S_{\mathrm{FP}}[\Phi_{\mu\nu},A_\mu,\varphi,\phi;\mathbf{g}]} \, \mathrm{Det}\big[ \hat{\Delta}_{(1)} \big] \, \mathrm{Det}\big[ \hat{\Delta}_{(0)} \big]\ ,
\end{equation}
with the total action put in diagonal form
\begin{equation}
 S_{\mathrm{FP}}[\Phi_{\mu\nu},A_\mu,\varphi,\phi;\mathbf{g}]=\frac12 S_{\mathrm{gra}}^{(\mathrm{I})}[\Phi_{\mu\nu};\mathbf{g}]+S_{\mathrm{vec}}[A_\mu;\mathbf{g}]+\frac{\mathfrak{m}^2}{2m^2} S_{\mathrm{sca}}[\varphi;\mathbf{g}]+\frac{\mathcal{X}}{2}\,S_{\mathrm{gra}}^{(\mathrm{II})}[\phi;\mathbf{g}]\ ,
\end{equation}
and recast only in terms of minimal kinetic operators. The scalar sector has some complicated prefactors, which may be eliminated through a simple field redefinition. It is now possible to path integrate over all the field species, producing the following functional determinants
\begin{equation} \label{3.3}
 Z[\mathbf{g}] = \mathrm{Det}^{-\nicefrac{1}{2}}_{\mathrm{\small TR}}\big[ \hat{\Delta}_{(2)} \big] \, \mathrm{Det}^{\nicefrac{1}{2}}\big[ \hat{\Delta}_{(1)} \big]\ .
\end{equation}
We denoted with the subscript ``TR" the fact that the functional determinant of the operator $\hat{\Delta}_{(2)}$ refers to traceless modes only. In order to exploit the calculations of \cite{Bastianelli:2023oca}, we have to manipulate the path integral to make the determinant correspond to \emph{full} rank-2 symmetric tensors. This can easily be achieved \cite{Vassilevich:2003xt}, to get as final outcome
\begin{equation}
 Z[\mathbf{g}] = \mathrm{Det}^{-\nicefrac{1}{2}}\big[ \hat{\Delta}_{(2)} \big] \, \mathrm{Det}^{\nicefrac{1}{2}}\big[ \hat{\Delta}_{(1)} \big] \, \mathrm{Det}^{\nicefrac{1}{2}}\big[ \hat{\Delta}_{(0)} \big]\ ,
\end{equation}
and the Euclidean one-loop effective action of linearized massive gravity finally reads
\begin{equation}
 \Gamma_{\mathrm{E}}[\mathbf{g}]=-\log Z[\mathbf{g}] =\frac12 \log\mathrm{Det}\big[ \hat{\Delta}_{(2)} \big]-\frac12 \log\mathrm{Det}\big[ \hat{\Delta}_{(1)} \big]-\frac12 \log\mathrm{Det}\big[ \hat{\Delta}_{(0)} \big]\ .
\end{equation}
This equation emphasizes that the Seeley-DeWitt coefficients of LMG can be determined by combining the individual coefficients related to the operators above. In particular, we may write
\begin{equation} \label{master}
\tr \big[a_j^{\mathrm{(tot)}}(x,D)\big]=\tr \big[a_j^{\mathrm{(gra)}}(x,D)\big]-\tr \big[a_j^{\mathrm{(vec)}}(x,D)\big]-\tr \big[a_j^{\mathrm{(sca)}}(x,D)\big]\ ,
\end{equation}
namely, we have reduced the problem of finding the HK expansion of the linearized massive gravity theory directly, which is in principle rather complicated due to the non-minimal character of the kinetic operator, to the one of expanding three simpler, massless,\footnote{Indeed, recall that the mass term is singled out in the Schwinger–DeWitt parametrization, and simply gets exponentiated in front of the effective action, acting as an IR cut-off. Therefore, the mass can be disregarded in the computation of the HK coefficients, as shown in appendix \ref{appB}.} theories. Most importantly, these are three theories for which such results are (at least partially) already available in the literature \cite{Bastianelli:2023oca, Groh:2011dw}. Each coefficient $a_j^{\mathrm{(gra/vec/sca)}}$ can be computed straightforwardly with the Schwinger-DeWitt technique \cite{Vassilevich:2003xt}, simply by studying the theory of a massless spin $2/1/0$ field coupled to on-shell gravity. For such theories, one has to consider the $m=0$ version of the minimal kinetic operator $\hat{\Delta}_{(2/1/0)}$, which is of the type
\begin{equation}
 \hat{\Delta}_{(2/1/0)} \sim -\nabla^2 + \mathcal{R}\ ,
\end{equation}
i.e. where the operator includes only specific non-minimal couplings $\mathcal{R}$ to the background curvature, in addition to the Laplacian term. \\

We report here the final result, referring the reader to appendix \ref{appB} for details regarding the extraction of the coefficients for each sector. Using the basis of geometric invariants up to order $\mathcal{O}(\mathcal{R}^3)$ on Einstein spaces (see appendix A of \cite{Bastianelli:2023oca}), we can express the heat-kernel coefficients as\footnote{We report only the dependence on the dimensionality of the background spacetime to stress the fact their value and, most importantly, their role vary with $D$. However, one should not forget that the curvature invariants $R(x), R_{\mu\nu\rho\sigma}(x), \dots$, which have to be inserted inside the spacetime integral, carry an $x$-dependence.} 
\begin{align}
\begin{split}
\tr \big[a_{0}(D)\big]&= \frac{(D+1)(D-2)}{2}\ , \label{a0}
\end{split}\\[.5em]
\begin{split}
\label{a1} \tr \big[a_{1}(D)\big]&= \frac{D^3-D^2-26 D-24}{12 D} \; R\ ,
\end{split}\\[.5em]
\begin{split}
\label{a2} \tr \big[a_{2}(D)\big]&=
\frac{5 D^4-7 D^3-248 D^2-596 D-1440}{720 D^2}\; R^2 +\frac{D^2-31 D+508}{360} \; R_{\mu\nu\rho\sigma}^{2}\ ,
\end{split}\\[.5em]
\tr \big[a_{3}(D)\big]&= \frac{35 D^5-77 D^4-2774 D^3-16532 D^2-135040D-483840}{90720 D^3}\;R^3 \nonumber\\[.3em]
&\phantom{=}+\frac{7 D^3-216 D^2+3219 D+11002}{15120 D}\;R\,R_{\mu\nu\rho\sigma}^{2} \label{a3} \\[.3em]
&\phantom{=}+ 
\frac{17 D^2-521 D-15658}{90720} \;R_{\mu\nu\rho\sigma}R^{\rho\sigma\alpha\beta}R_{\alpha\beta}{}^{\mu\nu}+\frac{D^2-37 D-1118}{3240}\; R_{\alpha\mu\nu\beta}R^{\mu\rho\sigma\nu}R_{\rho}{}^{\alpha\beta}{}_{\sigma}\ . \nonumber
\end{align}
The coefficients \eqref{a0}--\eqref{a3}, including the newly computed coefficient $a_3(D)$, allow for further investigations of the issue of one-loop divergences in massive gravity at the quantum level. The type of divergences can be seen explicitly by recasting the effective action as an expansion in powers of the Euclidean time:
\begin{equation} \label{divergences}
 \Gamma_{\mathrm{E}}[\mathbf{g}]= -\frac{1}{2}\int \frac{d^{D}x \sqrt{g}}{\left(4\pi \right)^{\frac{D}{2}}} \int_{0}^{+\infty}\frac{\diff\beta}{\beta^{\,1+\frac{D}{2}}} \, e^{-m^2\beta} \left[\tr(a_0) + \tr(a_1) \beta + \tr(a_2) \beta^2 + \tr(a_3) \beta^3+ \mathcal{O}(\beta^{4})\right]\ .
\end{equation} 
While the IR divergences are absent due to the presence of the mass, playing the role of a regulator, the UV divergences arise from the $\beta\to 0$ limit of the Euclidean time integration. To make contact with the logarithmic divergences usually encountered in QFT dimensional regularization \cite{Schwartz:2014sze}, one may use 
\begin{equation}
\int_{0}^{\infty}\frac{\diff\beta}{\beta^{\,1+\frac{D}{2}}}\beta^{\, p} \, e^{-m^{2}\beta}= (m^{2})^{\frac{D}{2}-p}\, \Gamma\left(p-\frac{D}{2}\right)\ ,
\end{equation}
and then expand the gamma function, taking $D=2p-2\epsilon$, in order to see the appearance of the usual $\frac{1}{\epsilon}$ pole as the leading divergent term. \\

In four spacetime dimensions, the different powers of $\beta$ give rise to the quartic, quadratic, and logarithmic divergences parametrized by $a_0, a_1,a_2$, respectively. In QFT dimensional
regularization only the logarithmic divergences are visible: they can be extracted from \eqref{a2}, and read
\begin{equation} \label{a2D4}
\left. \tr(a_{2}) \right|_{D=4}=-\frac{29}{48} \; R^2+\frac{10}{9} \; R_{\mu\nu\rho\sigma}^{2}\ .
\end{equation}
The latter numerical value for the one-loop four-dimensional logarithmic divergence of massive gravity coincides precisely with that calculated in \cite{Dilkes:2001av}. In $D$ dimensions, the $a_2$ coefficient can be extrapolated from the recent computations of \cite{Ferrero:2023xsf} and is correctly reproduced by our result \eqref{a2}. Finally, the coefficient $a_3$ gives rise to a finite term in the four-dimensional effective action: extracting from the HK expansion of the effective Lagrangian in \eqref{divergences} only the third order, we have
\begin{equation}
 \mathcal{L}^{(3)}_{\mathrm{E}}[\mathbf{g}]= -\frac{1}{32 \pi^2} \int_{0}^{+\infty} \diff\beta \; \mathrm{e}^{-m^2\beta} \, \left. \tr(a_{3}) \right|_{D=4}\ ,
\end{equation}
namely
\begin{equation} \label{derexp}
 \mathcal{L}^{(3)}_{\mathrm{E}}[\mathbf{g}]= -\frac{1}{32 \pi^2m^2} \left[ -\frac{4531}{18144} \; R^3 +\frac{2087}{6048} \; R\,R_{\mu\nu\rho\sigma}^{2}
-\frac{1747}{9072} \;R_{\mu\nu\rho\sigma}R^{\rho\sigma\alpha\beta}R_{\alpha\beta}{}^{\mu\nu}-\frac{125}{324}\; R_{\alpha\mu\nu\beta}R^{\mu\rho\sigma\nu}R_{\rho}{}^{\alpha\beta}{}_{\sigma} \right]\ .
\end{equation}
Note that, while for a massless theory the derivative expansion is not valid for getting finite terms of the effective action, as IR divergences arise, in our case \eqref{derexp} is a finite term in a consistent expansion. Setting the cosmological constant to zero, which implies $R=0$ on-shell, the remaining $R_{\mu\nu\rho\sigma}^{2}$ term in \eqref{a2D4} is a total derivative, and can be eliminated from the effective action. Therefore, the one-loop divergences are absent, and the leading term in the $4D$ effective Lagrangian reads in full glory 
\begin{equation}
 \boxed{\mathcal{L}_{\mathrm{E}}[\mathbf{g}]= \frac{1}{32 \pi^2m^2} \left[\frac{1747}{9072} \;R_{\mu\nu\rho\sigma}R^{\rho\sigma\alpha\beta}R_{\alpha\beta}{}^{\mu\nu}+\frac{125}{324}\; R_{\alpha\mu\nu\beta}R^{\mu\rho\sigma\nu}R_{\rho}{}^{\alpha\beta}{}_{\sigma} \right]}\ .
\end{equation}

Conversely, in $D=6$, the coefficient $a_3$ provides an additional divergence, the logarithmic one in that dimension:
\begin{equation} \label{D=6}
\left. \tr(a_{3}) \right|_{D=6} = -\frac{6893}{58320} \; R^3 +\frac{859}{3240} \; R\,R_{\mu\nu\rho\sigma}^{2}
-\frac{649}{3240} \;R_{\mu\nu\rho\sigma}R^{\rho\sigma\alpha\beta}R_{\alpha\beta}{}^{\mu\nu}-\frac{163}{405}\; R_{\alpha\mu\nu\beta}R^{\mu\rho\sigma\nu}R_{\rho}{}^{\alpha\beta}{}_{\sigma}\ .
\end{equation}
 More generally, the $a_3$ coefficient in $D$ dimensions has been partially evaluated in \cite{Fecit:2024jcv} with worldline techniques, and with the restriction to Ricci-flat spacetimes. Those results are in accordance with our computation \eqref{a3}, which, in addition, gives complete knowledge about the coefficient, going beyond the limitations of the wordline model there exploited \cite{Fecit:2023kah}. Taking $\Lambda=0$, the two surviving curvature invariants in \eqref{D=6} are generally independent of each other; nevertheless, in six dimensions there exists an integral relation that connects them \cite{vanNieuwenhuizen:1976vb} and the result can be expressed as follows
\begin{align}
\left. \tr(a_{3}) \right|_{D=6} = 
\frac{1}{1080} \;R_{\mu\nu\rho\sigma}R^{\rho\sigma\alpha\beta}R_{\alpha\beta}{}^{\mu\nu}\ ,
\end{align}
encoding the one-loop logarithmic divergences of massive gravity in six dimensions with vanishing cosmological constant. 

\section{Conclusions}
We have computed the first four heat-kernel coefficients for the Fierz-Pauli theory in a curved Einstein spacetime. While the coefficients for $n=0,1,2$ were already partially known in the literature, the fourth coefficient, denoted as $a_3(D)$, had not been fully determined. Indeed, previous calculations using first-quantized techniques required a Ricci-flat background to be valid at the quantum level. Interestingly, our results for the first four HK coefficients are in good agreement with the ones obtained via the worldline method, even beyond the expected regime of validity of the massive $\mathcal{N}=4$ spinning particle model. Specifically, we verified that terms proportional to the Ricci scalar, which were not reported in \cite{Fecit:2024jcv} and vanish upon projection to a Ricci-flat background, completely agree with \eqref{a1} and \eqref{a2}. The same holds for the $R\,R_{\mu\nu\rho\sigma}^{2}$ coefficient in \eqref{a3}, while the only exception is represented by a numerical prefactor in front of the $R^3$ invariant. For the sake of clarity, we report here the difference between the two results:
\begin{equation}
    \Delta\, \tr \big[a_3(D)\big]:=\tr \big[a^{\mathcal{N}=4}_3(D)\big]-\tr \big[a^{\mathrm{HK}}_3(D)\big]=\frac{(D+12)^3 }{432 D^3}R^3\ .
\end{equation}
This suggests the potential for refining the BRST quantization of the worldline model with a slight modification, which we leave for future investigation. \\
From a quantum field theory perspective, we have illustrated how gauge-fixing manipulations can significantly simplify the heat-kernel computations, bypassing the complexities of dealing directly with the non-minimal kinetic operators of the massive theory. Essential for the implementation of this method is the path integrals formulation of the Euclidean effective action and especially the (re)introduction of the gauge symmetry due to the redundant St\"uckelberg fields. While this procedure is technically cumberstone in generic $D$-dimensions, it has allowed us to leverage existing results in the literature directly. As a concrete application of our results, it would be interesting to compute the trace of the energy-momentum tensor in massive gravity and investigate its anomaly, as discussed in \cite{Ferrero:2023xsf}. Finally, it would be intriguing to place our methods and findings within the broader context of the growing interest in non-perturbative physics, particularly in the computation and application of resummed expressions \cite{Avramidi:2009quh, Edwards:2021cyp, Ahmadiniaz:2022bwy, Franchino-Vinas:2023wea} and the possible new effect
of pair creation in the presence of gravitational fields \cite{Ferreiro:2023jfs, Akhmedov:2024axn, Boasso:2024ryt}.

\section*{Acknowledgments}
It is a pleasure to thank F.~Ori for providing valuable insights on the heat-kernel computations and precious comments on the manuscript. We further thank F.~Bastianelli for his guidance during the unfolding of this work.

\appendix

\section{Seemingly singular case $m^2D=2R$} \label{appA}
We start from the action \eqref{2.5}, which is the FP theory for the tensor field $h_{\mu\nu}(x)$ on curved (Einstein) spacetime, with the introduction of a vector St\"uckelberg field $A_\mu(x)$, and where the newly-(re)introduced vector gauge symmetry has been fixed. We have
\begin{equation}
 S_{\mathrm{FP}}[h_{\mu\nu},A_\mu;\mathbf{g}]= \frac12 \int \diff^Dx \sqrt{g} \left[ h_{\mu\nu} \, \tilde{\Delta}^{\chi \;\mu\nu\rho\sigma}_{(\mathrm{II})} \, h_{\rho\sigma} 
+\mathcal{L}_{\mathrm{mix}}[A_\mu,h;\mathbf{g}] \right]\ ,
\end{equation}
where the minimal tensor operator can be read from \eqref{schifo} with $\chi=\nicefrac12$, $\chi'=1$, and where we have defined for later convenience the mixed Lagrangian
\begin{equation}
 \mathcal{L}_{\mathrm{mix}}[A_\mu,h;\mathbf{g}]=F_{\mu\nu}F^{\mu\nu}+\mathfrak{m}^2 A_\mu A^\mu-2m h \nabla_\mu A^\mu
\end{equation}
with the ``mass" $\mathfrak{m}^2$ reported in \eqref{mass}. \\

We proceed with the gauge-fixing manipulations avoiding the replacement \eqref{rep}, hence we will have to deal with the mixed tensor--vector term. We introduce the second St\"uckelberg field $\varphi(x)$ 
\begin{equation}
 A_\mu \to A_\mu +\frac{1}{m} \nabla_\mu \varphi\ ,
\end{equation}
which gives rise to new terms from the mixed Lagrangian
\begin{equation}
 \mathcal{L}_{\mathrm{mix}}[A_\mu,h;\mathbf{g}] \to \mathcal{L}_{\mathrm{mix}}[A_\mu,h,\varphi;\mathbf{g}]=\mathcal{L}_{\mathrm{mix}}[A_\mu,h;\mathbf{g}]+\frac{2\mathfrak{m}^2}{m}A^\mu\nabla_\mu \varphi +\frac{\mathfrak{m}^2}{m^2}\nabla_\mu \varphi\nabla^\mu\varphi-2h\nabla^2 \varphi\ ,
\end{equation}
We then fix the gauge symmetry \eqref{gauge2} by choosing
 \begin{equation} \label{gauge-fix3}
 \nabla^\mu A_\mu+\frac{\mathfrak{m}^2}{2m}\varphi+\frac{m}{2}h=0\ ,
\end{equation}
which, on the one hand, solves the tensor--vector mixing and makes the vector/scalar minimal operators appear, but, on the other hand, complicates further the tensor operator and leaves the newly introduced trace--scalar mixing: the action now reads
\begin{align}
S_{\mathrm{FP}}[h_{\mu\nu},A_\mu,\varphi;\mathbf{g}]= \frac12 \int \diff^Dx \sqrt{g} \left[ h_{\mu\nu} \, \tilde{\Delta}^{\chi \;\mu\nu\rho\sigma}_{(\mathrm{II})} \, h_{\rho\sigma} + 2 A_{\mu} \, \hat{\Delta}^{\mu\nu}_{(1)} \, A_{\nu}+\frac{\mathfrak{m}^2}{m^2} \, \varphi \, \hat{\Delta}_{(0)} \, \varphi -2h\nabla^2 \varphi+\mathfrak{m}^2h\varphi \right]\ ,
\end{align}
where the spin 2 operator has been defined in \eqref{schifo} with $\chi=\nicefrac12$ and with now, differently from before, $\chi'=\nicefrac12$ too. It is once again convenient to decompose $h_{\mu\nu}$ into traceless and scalar parts as in \eqref{decompose}, so as to make evident that the only difficulties come from the mixing in the scalar sector of the theory:
\begin{equation} \label{A.7}
 S_{\mathrm{FP}}[\Phi_{\mu\nu},A_\mu,\varphi,\phi;\mathbf{g}]=\frac12 S_{\mathrm{gra}}^{(\mathrm{I})}[\Phi_{\mu\nu};\mathbf{g}]+S_{\mathrm{vec}}[A_\mu;\mathbf{g}]+ S_{\mathrm{sca/mix}}[\varphi,\phi;\mathbf{g}]\ ,
\end{equation}
where the minimal and diagonal spin 2 action is reported in \eqref{2.23}, the minimal and diagonal vector action can be read off from \eqref{2.19}, and where the scalar sector explicitly reads
\begin{equation}
 S_{\mathrm{sca/mix}}[\varphi,\phi;\mathbf{g}]= \int \diff^Dx \sqrt{g} \left[ \frac{\mathfrak{m}^2}{2m^2} \, \varphi \, \hat{\Delta}_{(0)} \, \varphi-\frac{D-2}{4D} \, \phi \, \hat{\Delta}_{(0)} \, \phi +\varphi \, \hat{\Delta}_{(0)} \, \phi \right]\ .
\end{equation}
The latter equation can be expressed in quadratic form as
\begin{equation}
 S_{\mathrm{sca/mix}}[\mathbf{\Psi};\mathbf{g}]\int \diff^Dx \sqrt{g} \; \Psi^A \, M_{AB} \, \Psi^B\ ,
\end{equation}
by means of
\begin{equation}
 \Psi^A := \left(\begin{array}{c} \phi \\ \varphi \end{array} \right)\ , \quad M_{AB}:=\left( \begin{array}{cc} \tfrac{2-D}{4D} & 0 \\ 1 & \frac{\mathfrak{m}^2}{2m^2} \end{array} \right) \, \hat{\Delta}_{(0)}\ .
\end{equation}
Note that path integrating the latter action simply produces
\begin{equation}
 \int [\mathcal{D}\mathbf{\Psi}] \, e^{ S_{\mathrm{sca/mix}}[\mathbf{\Psi};\mathbf{g}]} = \mathrm{Det}^{-\nicefrac{1}{2}}\big[ M \big]\ ,
\end{equation}
up to a rescaling of the fields. Therefore, since the scalar sector is decoupled from the tensor and vector one, we can proceed by path integrating them separately with the action \eqref{A.7}, yielding
\begin{equation}
 Z_{\mathrm{sca/mix}}[\mathbf{g}]:= \int [\mathcal{D}\varphi] \, [\mathcal{D}\phi] \, e^{ S_{\mathrm{sca/mix}}[\varphi,\phi;\mathbf{g}]} = \mathrm{Det}^{-1}\big[ \hat{\Delta}_{(0)} \big]\ ,
\end{equation}
where the last equality is valid up to some prefactors, which can be absorbed inside the path integral normalization or by a field redefinition. Therefore, inserting the scalar result in the full path integration 
\begin{equation}
 Z[\mathbf{g}] = Z_{\mathrm{sca/mix}}[\mathbf{g}] \int [\mathcal{D}\mathbf{\Phi}] \, [\mathcal{D}\mathbf{A}] \, e^{\tfrac12 S_{\mathrm{gra}}^{(\mathrm{I})}[\Phi_{\mu\nu};\mathbf{g}]+S_{\mathrm{vec}}[A_\mu;\mathbf{g}]} \, \mathrm{Det}\big[ \hat{\Delta}_{(1)} \big] \, \mathrm{Det}\big[ \hat{\Delta}_{(0)} \big]\ ,
\end{equation}
where we included also the required $\Phi\Pi$ determinants, we get the final result
\begin{equation}
 Z[\mathbf{g}] = \mathrm{Det}^{-\nicefrac{1}{2}}_{\mathrm{\small TR}}\big[ \hat{\Delta}_{(2)} \big] \, \mathrm{Det}^{\nicefrac{1}{2}}\big[ \hat{\Delta}_{(1)} \big]\ ,
\end{equation}
which coincides with equation \eqref{3.3}, obtained by using the replacement \eqref{rep}, which has been avoided in this appendix.

\section{Heat-kernel coefficients for each sector} \label{appB}

\subsection{General formulae}
Consider a generic elliptic second-order (minimal) differential operator on a $D$-dimensional Euclidean manifold
\begin{equation}\label{DiffOp}
\Delta= -\nabla^2-V\ ,
\end{equation}
where $\nabla^2$ is the Laplacian, $V$ is a matrix potential, which may include a mass term. We denote by $\Omega_{\mu\nu}(x)$ the field-strength,
\begin{equation} \label{B.2}
\Omega_{\mu\nu} = [\nabla_\mu, \nabla_\nu]\ ,
\end{equation}
acting on scalars, vectors, and symmetric tensors. Its form will be specified from the commutator acting on each of the field spaces. \\

It is possible to express the heat-kernel coefficients $a_j(x,D)$ associated with the operator \eqref{DiffOp} -- defined through the heat-kernel expansion
\begin{equation}\label{HDMS}
K(x,x;\beta)\sim (4\pi \beta)^{-\nicefrac{D}{2}}\sum_{j=0}^{\infty} \beta^j a_j(x,D)\ ,
\end{equation}
where $K(x,x;\beta)$ is the diagonal heat-kernel -- only in terms of the metric and of the gauge invariants of the manifold, namely in terms of $g_{\mu\nu}$, the connection $\Omega_{\mu\nu}$ and the potential term $V$. The SdW coefficients have been expressed in these general terms for a generic theory up to $a_5(x,D)$ \cite{Avramidi2015}; here we are interested in reaching the third order, namely we want to compute said coefficients up to $a_3(x,D)$. The general formulae can be written in terms of
\begin{equation}\label{HKalphabeta}
a_j(x,D) \equiv \alpha_j(x,D) + \gamma_j(x,D)\ ,
\end{equation}
where $\alpha_j(x,D)$ comes from considering an exponentiated form of the heat-kernel series,
\begin{equation}\label{HKconnected}
\Tr{\left[\sum_{j=0}^{\infty} \beta^j a_j(x,D) \right]}\equiv \Tr{\left[\exp{\left(\sum_{j=1}^{\infty} \beta^j \alpha_j(x,D)\right)}\right]}\ ,
\end{equation}
while the remainders $\gamma_j(x,D)$ read
\begin{equation}\label{BetaValues}
\gamma_0=\gamma_1 = 0\ ,\quad
\gamma_2 = \frac{1}{2}\alpha_1^2\ ,\quad
\gamma_3 = \frac{1}{6}\alpha_1^3+\alpha_1\alpha_2\ .
\end{equation}
The explicit expressions are given by \cite{Gilkey:1975iq, Bastianelli:2000hi}
\begin{align}
\begin{split}
\label{HK0} \alpha_0(D) &= \mathbbm{1}
\end{split}\\[.5em]
\begin{split}
\label{HK1} \alpha_1(x,D) &= \frac{1}{6}R\mathbbm{1} + V
\end{split}\\[.5em]
\begin{split}
\label{HK2} \alpha_2(x,D) &= \frac{1}{6}\nabla^2\left(\frac{1}{5}R\mathbbm{1}+V\right)+ \frac{1}{180}\left(R_{\mu\nu\rho\sigma}^2-R_{\mu\nu}^2\right)\mathbbm{1} + \frac{1}{12}\Omega_{\mu\nu}^2
\end{split}\\[.5em]
\alpha_3(x,D) &= \frac{1}{7!}\left[18\nabla^4 R + 17(\nabla_\mu R)^2-2(\nabla_\mu R_{\nu\sigma})^2 - 4\nabla_\mu R_{\nu\sigma}\nabla^\nu R^{\mu\sigma}+9(\nabla_\alpha R_{\mu\nu\rho\sigma})^2 -8R_{\mu\nu}\nabla^2 R^{\mu\nu}\right. \nonumber \\
&\phantom{=}
+12R^{\mu\nu}\nabla_\mu \nabla_\nu R
+12R_{\mu\nu\rho\sigma}\nabla^2R^{\mu\nu\rho\sigma} 
+ \frac{8}{9}R_{\mu}{}^{\nu}R_{\nu}{}^{\sigma}R_{\sigma}{}^{\mu} - \frac{8}{3}R_{\mu\nu}R_{\rho\sigma}R^{\mu\rho\nu\sigma} \nonumber
\\
&\phantom{=}\left.-\frac{16}{3}R_{\mu\nu}R^\mu{}_{\rho\sigma\tau}R^{\nu\rho\sigma\tau}+\frac{44}{9}R_{\mu\nu}{}^{\rho\sigma}R_{\rho\sigma}{}^{\alpha\beta}R_{\alpha\beta}{}^{\mu\nu}+\frac{80}{9}R_{\mu\nu\rho\sigma}R^{\mu\alpha\rho\beta}R^{\nu}{}_{\alpha}{}^{\sigma}{}_{\beta}\right]\mathbbm{1} \nonumber \\
&\phantom{=}+\frac{2}{6!}\left[8(\nabla_\mu \Omega_{\nu\sigma})^2+2(\nabla^\mu \Omega_{\mu\nu})^2 + 12\Omega_{\mu\nu}\nabla^2\Omega^{\mu\nu}-12\Omega_{\mu}{}^{\nu}\Omega_{\nu}{}^{\sigma}\Omega_{\sigma}{}^{\mu}+6R_{\mu\nu\rho\sigma}\Omega^{\mu\nu}\Omega^{\rho\sigma}-4R_{\mu\nu}\Omega^{\mu\sigma}\Omega^\nu{}_{\sigma}\right. \nonumber \\
&\phantom{=}\left. +6\nabla^4 V+30(\nabla_\mu V)^2+4R_{\mu\nu}\nabla^\mu\nabla^\nu V+12\nabla_\mu R\nabla^\mu V\right]\ . \label{HK3}
\end{align}
Therefore, given the explicit form of a second-order (minimal) differential operator $\hat{\Delta}_{(i)}$, we can compute its HK coefficients by correctly identifying $\mathbbm{1}$, which represent the identity on the field space, together with $V$ and $\Omega_{\mu\nu}$, through a comparison with \eqref{DiffOp}--\eqref{B.2}. We will proceed with this identification for each sector of the Fierz-Pauly theory on curved Einstein background, i.e. for the operators $\hat{\Delta}_{(2/1/0)}$, in the remainder of this appendix. A check is performed for each sector with results reported in the literature. \\

Let us comment on the fact that, even though each of the operators considered in this work contains a mass term, $m$ will be ignored in identifying the desired gauge invariants $V_{(i)}$. Indeed, the one-loop effective action can be recast by keeping the mass outside of the HK expansion as an exponential factor, highlighting that the mass term acts as a cut-off for infrared divergences.

\subsection{Scalar sector}
The differential operator of St\"uckelberg scalar is given by \eqref{Delta0}
\begin{equation}
\hat{\Delta}_{(0)}=-\nabla^2 +m^2-\frac{2R}{D}\ ,
\end{equation}
which is an elliptic second-order (minimal) differential operator, thus it can be treated within the heat-kernel expansion. The SdW coefficients \eqref{HK0}--\eqref{HK3} can be computed by identifying the explicit formulae for the gauge invariants previously introduced, among which $\Omega_{\mu\nu}$ is given by \eqref{B.2}
\begin{equation}
 [\nabla_\mu, \nabla_\nu] \, \phi =0\ ,
\end{equation}
acting on a test scalar field. We get
\begin{equation}
\begin{cases}
\mathbbm{1}_{(0)}\ \leftrightarrow\ 1 \\[.3em]
V_{(0)}\ \leftrightarrow\ \dfrac{2R}{D} \\[.5em]
\Omega^{(0)}_{\mu\nu}\ \leftrightarrow\ 0\ ,
\end{cases}
\end{equation}
where the mass has been neglected, for the reasons previously commented on. The scalar coefficients follow by direct computation and read
\begin{align}
\begin{split}
\tr \big[a_{0}^{(\mathrm{sca})}(D)\big]&= 1
\end{split}\\[.5em]
\begin{split}
\tr \big[a_{1}^{(\mathrm{sca})}(D)\big]&= \frac{D+12}{6D}\; R
\end{split}\\[.5em] 
\begin{split}
\tr \big[a_{2}^{(\mathrm{sca})}(D)\big]&=\frac{5 D^2+118 D+720}{360 D^2}\; R^2 + \frac{1}{180}\; R_{\mu\nu\rho\sigma}^{2} 
\end{split}\\[.5em]
\begin{split}
\tr \big[a_{3}^{(\mathrm{sca})}(D)\big]&=\frac{70 D^3+2499 D^2+29980 D+120960}{22680 D^3}\; R^3 +\frac{7 D+85}{7560 D}\;R\, R_{\mu\nu\alpha\beta}^{2} \\[.3em] 
&\phantom{=}+\frac{17}{45360}\;R_{\mu\nu\rho\sigma}R^{\rho\sigma\alpha\beta}R_{\alpha\beta}{}^{\mu\nu} +\frac{1}{1620}\;R_{\alpha\mu\nu\beta}R^{\mu\rho\sigma\nu}R_{\rho}{}^{\alpha\beta}{}_{\sigma}\ .
\end{split}
\end{align}
These coefficients agree with known results in the literature, see for instance \cite{Groh:2011dw, Ferrero:2023xsf}. Once plugged inside \eqref{master}, together with the vector and tensor coefficients, they give the final result.

\subsection{Vector sector}
The differential operator of St\"uckelberg vector has been reduced through a gauge-fixing manipulation to \eqref{Delta1}
\begin{equation} 
 \hat{\Delta}^{\mu\nu}_{(1)}=\left(-\nabla^2 +m^2-\frac{R}{D}\right) \, g^{\mu\nu}\ ,
\end{equation}
which is an elliptic second-order (minimal) differential operator. Recalling that the commutator acts on a test vector field as
\begin{equation}\label{GhostComm}
[\nabla_\mu, \nabla_\nu] \, A^\rho=R_{\mu\nu}{}^{\rho}{}_{\sigma} \, A^\sigma\ ,
\end{equation}
we conclude that the substitutions to be performed in the general formulae are
\begin{equation}\label{GhostSost}
\begin{cases}
\mathbbm{1}_{(1)}\ \leftrightarrow\ \delta^\mu_\nu\\[.3em]
V_{(1)}\ \leftrightarrow\ \dfrac{R}{D}\delta^\mu_\nu\\[.5em]
(\Omega^{(1)}_{\mu\nu})^\rho_{\ \sigma} \ \leftrightarrow\ R_{\mu\nu}{}^{\rho}{}_{\sigma}\ .
\end{cases}
\end{equation}
The resulting computations (see appendix C of \cite{Bastianelli:2023oca} for details) produce
\begin{align}
\begin{split}
\label{a0gh} \tr \big[a_{0}^{(\mathrm{vec})}(D)\big]&= D
\end{split}\\[.5em]
\begin{split}
\label{a1gh} \tr \big[a_{1}^{(\mathrm{vec})}(D)\big]&= \frac{D+6}{6}\; R
\end{split}\\[.5em]
\begin{split}
\label{a2gh} \tr \big[a_{2}^{(\mathrm{vec})}(D)\big]&=\frac{5 D^2+58 D+180}{360 D}\; R^2 + \frac{D-15}{180}\; R_{\mu\nu\rho\sigma}^{2}
\end{split}\\[.5em]
\begin{split}
\label{a3gh} \tr \big[a_{3}^{(\mathrm{vec})}(D)\big]&=\frac{35 D^3+588 D^2+3512 D+7560}{45360 D^2}\; R^3 +\frac{7 D^2-62 D-714}{7560 D}\;R\, R_{\mu\nu\alpha\beta}^{2} \\[.3em]
&\phantom{=}+\frac{17 D-252}{45360}\;R_{\mu\nu\rho\sigma}R^{\rho\sigma\alpha\beta}R_{\alpha\beta}{}^{\mu\nu} +\frac{D-18}{1620}\;R_{\alpha\mu\nu\beta}R^{\mu\rho\sigma\nu}R_{\rho}{}^{\alpha\beta}{}_{\sigma}\ .
\end{split}
\end{align}
These coefficients agree with known results in the literature, see for instance \cite{Groh:2011dw, Ferrero:2023xsf}. Once plugged inside \eqref{master}, together with the scalar and tensor coefficients, they give the final result.

\subsection{Tensor sector}
The tensor operator has been reduced through gauge-fixing manipulations to \eqref{DeltaI}
\begin{equation}
 \hat{\Delta}^{\mu\nu\rho\sigma}_{(2)}=\left(-\nabla^2 +m^2\right) \, g^{\mu(\rho}g^{\sigma)\nu} +2R^{\mu(\rho\sigma)\nu}\ ,
\end{equation}
which is an elliptic second-order (minimal) differential operator. Recalling that the commutator acting on the spin 2 field is
\begin{equation}\label{GravitonComm}
[\nabla_\mu, \nabla_\nu] \, h_{\rho\sigma}=R_{\rho\sigma}{}^{\alpha\beta}{}_{\mu\nu} \, h_{\alpha\beta}\ ,
\end{equation}
where
\begin{equation}\label{SuperRiemann}
R_{\rho\sigma}{}^{\alpha\beta}{}_{\mu\nu}\equiv \frac{1}{2}\left(\delta_\rho^\alpha R_{\sigma}{}^{\beta}{}_{\mu\nu} + \delta_\rho^\beta R_{\sigma}{}^{\alpha}{}_{\mu\nu} +\delta_\sigma^\alpha R_{\rho}{}^{\beta}{}_{\mu\nu} + \delta_\sigma^\beta R_{\rho}{}^{\alpha}{}_{\mu\nu} \right)\ ,
\end{equation}
the explicit form of the invariants $\mathbbm{1}$, $V$ and $\Omega_{\mu\nu}$ is
\begin{equation}\label{GravitonSost}
\begin{cases}
\mathbbm{1}_{(2)}\ \leftrightarrow\ \delta_{\mu\nu}^{\ \ \ \alpha\beta}\\
V_{(2)}\ \leftrightarrow\ \mathcal{V}_{\mu\nu}{}^{\alpha\beta}\\
(\Omega^{(2)}_{\mu\nu})_{\rho\sigma}{}^{\alpha\beta} \ \leftrightarrow\ R_{\rho\sigma}{}^{\alpha\beta}{}_{\mu\nu}\ ,
\end{cases}
\end{equation}
where
\begin{align}
\label{SuperDelta} \delta_{\mu\nu}{}^{\alpha\beta} &\equiv \frac{1}{2}\left(\delta_\mu^\alpha\delta_\nu^\beta+\delta_\mu^\beta\delta_\nu^\alpha\right)\ ,\\[.5em]
\label{VGraviton} \mathcal{V}_{\mu\nu}{}^{\alpha\beta} &\equiv R_{\mu}{}^{\alpha}{}_{\nu}{}^{\beta} + R_{\mu}{}^{\beta}{}_{\nu}{}^{\alpha}\ .
\end{align}
The resulting computations are tedious but straightforward (see appendix C of \cite{Bastianelli:2023oca} for details), and the tensor coefficients read
\begin{align}
\begin{split}
\label{a0gr} \tr \big[a_{0}^{(\mathrm{gra})}(D)\big]&= \frac{D (D+1)}{2} 
\end{split}\\[.5em]
\begin{split}
\label{a1gr} \tr \big[a_{1}^{(\mathrm{gra})}(D)\big]&= \frac{\left(D^2+D-12\right)}{12} \; R
\end{split}\\[.5em]
\begin{split}
\label{a2gr} \tr \big[a_{2}^{(\mathrm{gra})}(D)\big]&=\frac{\left(5 D^2+3 D-122\right)}{720} \; R^2 +\frac{\left(D^2-29 D+480\right)}{360} \; R_{\mu\nu\rho\sigma}^{2}
\end{split}\\[.5em]
\label{a3gr} \tr \big[a_{3}^{(\mathrm{gra})}(D)\big]&= \frac{35 D^3-7 D^2-1318 D+488}{90720 D}\;R^3+ \frac{7 D^3-202 D^2+3109 D+9744}{15120 D}\;R\, R_{\mu\nu\alpha\beta}^{2} \\[.3em]
&\phantom{=}+\frac{17 D^2-487 D-16128}{90720}\;R_{\mu\nu\rho\sigma}R^{\rho\sigma\alpha\beta}R_{\alpha\beta}{}^{\mu\nu}+\frac{D^2-35 D-1152}{3240}\;R_{\alpha\mu\nu\beta}R^{\mu\rho\sigma\nu}R_{\rho}{}^{\alpha\beta}{}_{\sigma}\ . \nonumber
\end{align}
These coefficients agree with known results in the literature, see for instance \cite{Groh:2011dw, Ferrero:2023xsf}. Once plugged inside \eqref{master}, together with the scalar and vector coefficients, they give the final result.


\addcontentsline{toc}{section}{References}
\printbibliography


\end{document}